\documentclass[epj,referee]{svjour}
\usepackage{graphics}
\newcommand{\lambdabar}{{\mkern0.75mu\mathchar '26\mkern -9.75mu\lambda}}
\begin{document}
\title{The effect of the  neutron and proton numbers ratio in colliding nuclei
 at formation of the evaporation residues in the $^{34}$S+$^{208}$Pb
and  $^{36}$S+$^{206}$Pb reactions.}
\author{A.K. Nasirov\inst{1,2}, B.M. Kayumov\inst{2},G.~Mandaglio\inst{3,4},
G. Giardina\inst{5}, K. Kim\inst{6},  Y. Kim\inst{6}}
\offprints{Avazbek Nasirov}          
\mail{Joint Institute for Nuclear Research, Joliot-Curie 6, 141980
Dubna,
 Russia}
\institute{BLTP, Joint Institute for Nuclear Research, Joliot-Curie
6, 141980 Dubna, Russia \and Institute of Nuclear Physics, Ulugbek,
100214, Tashkent, Uzbekistan
\and Dipartimento di Scienze Chimiche, Biologiche, Farmaceutiche ed Ambientali, University of Messina, Messina, Italy
\and INFN Sezione di Catania, Catania, Italy
\and Dipartimento di Scienze Matematiche e Informatiche, Scienze Fisiche e Scienze della Terra, University of Messina, Messina, Italy
 \and Rare Isotope Science Project,
Institute for Basic Science, Daejeon 305-811, Republic of Korea}
\date{Received: date / Revised version: date}
%
\abstract{
The difference between observed cross sections of the evaporation residues (ER)
 of the $^{34}$S+$^{208}$Pb and $^{36}$S+$^{206}$Pb reactions
formed in the 2n and 3n channels has been explained by two reasons
related with the entrance channel characteristics of these reactions.
The first reason is that the capture cross section of the latter reaction is larger
than the one of the $^{34}$S+$^{208}$Pb reaction since the nucleus-nucleus potential
 is more attractive in the $^{36}$S+$^{206}$Pb reaction due to two more neutrons in
isotope $^{36}$S. The second reason is the difference
in the heights of the intrinsic fusion barrier $B^*_{\rm fus}$ appearing on the fusion
 trajectory by nucleon transfer between
nuclei of the DNS formed after the capture. The value of $B^*_{\rm fus}$ calculated
 for the $^{34}$S+$^{208}$Pb reaction is higher than the one obtained for the
  $^{36}$S+$^{206}$Pb  reaction. This fact has been caused by the difference between
   the $N/Z$-ratios in the light fragments of the DNS formed during the capture in
   these reactions.
The  $N/Z$-ratio has been found by solution of the transport master equations
 for the proton and neutron distributions between fragments of the DNS
 formed at capture with the different initial neutron numbers
 $N=18$ and $N=20$ for the reactions with the $^{34}$S and $^{36}$S, respectively.
}
\titlerunning{The role of the N/Z-ratio in colliding nuclei in formation}
\authorrunning{A.K. Nasirov et al.}
\maketitle

\section{Introduction}
\label{int}
The complete fusion of two colliding nuclei and de-excita-tion
of the formed excited compound nucleus (CN) are topics of a great
interest in  physics of nuclear reaction. The de-excitation of the CN  through
the evaporation of neutrons can lead to the synthesis of
the heaviest elements (with the charge numbers $Z$ = 108--118).
A significant difficulty in producing the heaviest elements in
fusion-evaporation reactions is very small cross sections, which
can be attributed to the hindrance at the formation of the CN and/or
to its instability against fission.
The high accuracy of measurements of the ER cross section
stimulates studies of the peculiarities of the reaction mechanism
in experiments producing transuranic elements.
For this aim, in Ref.\cite{KNHAB12}, the authors compared
two reactions with the different isotopes of S and Pb leading to the same nucleus $^{242}$Cf.
 The measured cross section for the 2n channel of $^{36}$S+$^{206}$Pb was
approximately 25 times larger than that of $^{34}$S+$^{208}$Pb.
In the case of the 3n channel,  approximately ten times larger cross
section was measured for the $^{36}$S+$^{206}$Pb reaction compared to the 
$^{34}$S+$^{208}$Pb reaction.
These values were obtained close to the maxima of the 2n and 3n
cross sections.
In Ref.\cite{KNHAB15}, the experimental data of the yield of fission
products have been obtained to establish the main reasons causing
the difference in the results of the evaporation residue cross sections
in the above mentioned reactions. The excitation
function of fission for the $^{36}$S+$^{206}$Pb reaction is higher than the one
of the $^{34}$S+$^{208}$Pb reaction if they are compared as  functions
of the CN excitation energy.
 In Ref. \cite{KNHAB12}, the authors noted
that the difference of only two neutrons in projectile and target
nuclei has such a strong influence on the fusion probability.
This amazing result requests its further explanation.

The first attempt was made by the authors of the experiments in Ref.\cite{KNHAB12}
by the description of the fission excitation functions using the coupled-channels calculations
(code CCFULL \cite{CCFUL}) with coupling to vibrational states; the evaporation residue cross sections
 were estimated by the statistical code HIVAP \cite{HIVAP}, normalizing the capture cross section
 calculated by the HIVAP to
the measured capture cross sections. Regardless of this procedure, the authors modified the parameters
of the HIVAP code  for the survival probability of $^{242}$Cf$^*$ such that the measured
ER cross sections of the $^{36}$S+$^{206}$Pb reaction were reproduced at $E^*$= 25.5
and 33.1 MeV for the 2n and 3n channels, respectively. This action allowed authors to reproduce
the ER cross sections of the $^{36}$S+$^{206}$Pb reaction, but at the use of the same parameters
 leads to overestimation of the measured data of the $^{34}$S+$^{208}$Pb reaction.
 Therefore, authors have concluded
that the latter reaction exhibits a significant hindrance of fusion relative to the $^{36}$S+$^{206}$Pb
 reaction.
Another attempt to clarify the nature of the hindrance to complete fusion in the case
of the $^{34}$S+$^{208}$Pb reaction has been done in Ref.\cite{KNHAB15}.
The authors of Ref.\cite{KNHAB15} have used new fission measurements and existing evaporation residue and fission excitation
function data for reactions forming Cf isotopes to investigate the dependence of the
quasifission probability and characteristics of fission products on the properties
of the entrance channels.
The calculations made by the use of the coupled-channels code CCFULL reproduce
the measured capture cross sections for both reactions.
Assuming no quasifission ($P_{CN}$=1), the statistical model calculations of the fission
 survival probability $W_{CN}$ are able to reproduce the measured $x$n ER cross sections
  for the $^{36}$S+$^{206}$Pb reaction.
  However, the calculations made by the use of the same parameters have led to the
  overestimation of  the experimental $x$n ER cross sections for the $^{34}$S+$^{208}$Pb
   reaction. The agreement with the experimental data of the last reaction
can be reached by introducing a strong fusion hindrance for it relative to
the  $^{36}$S+$^{206}$Pb reaction (with an angular momentum
averaged hindrance $P_{CN}$=0.1). This procedure must be associated with a large
 quasifission probability for $^{34}$S+$^{208}$Pb.
 The analysis of mass and angle distributions of the fusion-fission products
of these two reactions did not reveal a significant difference between them showing the hindrance
in fusion.
The authors concluded that the strongly hindered ER yield for the $^{34}$S+$^{208}$Pb reaction
compared with the $^{36}$S+$^{206}$Pb reaction indicates that the quasifission competition is
weaker in the $^{36}$S+$^{206}$Pb reaction. This must be attributed to the different nuclear
structures of the reaction partners in the two reactions and  closer matching of $N/Z$ ratios
in the latter reaction, as found in Ca + Pb reactions \cite{SimenelPLB12}.
The authors could not explain the appearance of the hindrance to fusion causing the value $P_{CN}$=0.1.
One of the reasons of difficulties in study of hindrance to fusion is difference
 in the theoretical and experimental views to the capture events.

In this work we try to establish reasons causing difference at the CN formation
 in the two reactions under discussions by the theoretical analysis
 of the formation of dinuclear system (DNS) after capture and its
 transformation into the CN.
In Section \ref{views} the different views in estimation of the
capture cross section are shortly discussed.
 The methods of calculations are briefly presented in Section \ref{model}
 and results of the capture and fusion cross sections are discussed
 in Section \ref{results}.

\section{Different views to the definition of capture}
\label{views}

 In the deep-inelastic collisions, the full momentum transfer
 of the relative motion does not take place and interaction time of the
  colliding nuclei is relatively
  shorter than in the case of capture reactions which request the full momentum transfer.
 The main difference between  deep-inelastic collision and capture events,
  which can be observed in experiment, is a value of the total kinetic energy of the reaction products.
  The total kinetic energy of the products formed in the capture reaction are fully damped
  and its value is significantly lower than the initial collision energy $E_{\rm c.m.}$,
 while the total kinetic energy of the deep-inelastic collisions products is not fully
 damped and its value is close to the $E_{\rm c.m.}$.
 The DNS formed as a result of the capture of the colliding nuclei
  can evolve to one of states of the heated and rotating compound nucleus
 (complete fusion) or it breaks down forming two
fragments  (quasifission) without reaching the saddle point of CN.

  The mass and charge distributions of the  deep-inelastic collision and
 capture events can widely overlap. This overlaps of the mass and charge distributions have been
  discussed
 in Ref. \cite{MPHL05} for the case of $^{48}$Ca+$^{208}$Pb reaction.
 The main conclusion from this short comment is that capture events are presented as the yield
 of the projectile- and target-like products with the total kinetic energy significantly
 lower than the initial collision energies.
  The total kinetic energy of the  products formed in the capture reaction
  are around their Coulomb barriers in the exit channels since the amount of the kinetic energy
  of the relative motion above the Coulomb barrier is dissipated, {\it i.e.} the full momentum
  of the relative motion occurs. The difference between the total kinetic energies 
  of the products formed in 
  the deep-inelastic collision and capture events depends on the projectile-nucleus
  energy, orbital angular momentum of collision, mass and charge
 numbers of the colliding nuclei.
   Unfortunately, there is not so many experimental and theoretical
 studies devoted to the important problem which allows us to draw interesting
 conclusions about reaction mechanism of the heavy-ion collisions at the energies
 near the Coulomb barrier. Since the events producing  projectile-like and target-like binary products
 are considered as the deep-inelastic collision events only.
Therefore, the separation of the capture events
 producing projectile-like and target-like binary products from the deep-inelastic collision events
 requests detailed analysis of the experimental data and developing corresponded
 theoretical methods.
 Nevertheless, there are papers where the authors have studied  the properties
 of the reaction products by the analysis of their total kinetic energies.
 For example, in Fig. 3 of Ref.\cite{Mohanto2018}, the yield of the binary
 products with the  mass numbers in the range $M_1=40$---56 of the
 $^{50}$Cr+$^{208}$Pb reaction and having total kinetic energy around 235 MeV
  are shown as to be belonged to quasielastic and the ones having
  the total kinetic energy around 160 MeV are marked as the products
  of the deep-inelastic collisions. All of the
 products with the mass numbers in the range $M_1=57$---82 are indicated as ones
 of the fast quasifission process. According to our point of view, among
 the products marked the deep-inelastic collisions there are events of
 the quasifission  having the total kinetic energy approximately
  in the range 150---170 MeV. More detailed analysis should be performed
  in our future research devoted to this topic.
   The yield of the projectile- and target-like products of the capture reactions
 is responsible for the decrease of the events going to the complete fusion and this mechanism can be considered as hindrance to complete fusion which is not studied by experimentalists.

 The correct estimated quasifission cross section $\sigma_{\rm qf}$ contains a contribution
  of the yield of projectile- ($\sigma_{PLqf}$) and target-like ($\sigma_{TLqf}$ products
   together  with asymmetric ($\sigma_{asymqf}$) and symmetric ($\sigma_{symqf}$) fragments of
    the DNS decay:
  \begin{equation}
  \sigma_{\rm qf}=\sigma_{PLqf}+\sigma_{TLqf}+\sigma_{asymqf}+\sigma_{symqf}.
  \label{qf}
  \end{equation}
  Usually the contributions $\sigma_{PLqf}$ and $\sigma_{TLqf}$ to quasifission are
  not considered at the estimation of the capture though the strong yield of these products
  decreases the amount of events leading to complete fusion.

 The authors of Refs. \cite{KNHAB12} and \cite{KNHAB15} have concentrated their attention
 to the mass region $0.25<M_R<0.75$ in analysis of the mass-angle distribution of the fusion-fission products.
 Then the fusion-fission cross sections were considered as capture cross section since
 the authors had assumed that there was no hindrance to the complete fusion  in
 the $^{36}$S+$^{206}$Pb reaction. The mass region of the quasifission products, which overlaps
 with the one of the projectile-like and target-like products of deep-inelastic collision, is
 around $M_R=0.15$. The authors of Refs. \cite{KNHAB12} and \cite{KNHAB15} did not study this
  mass region of the reaction products.

    Unfortunately,  often these quasifission products are considered as products of
  the deep-inelastic collisions and their contribution is not included in the
  capture cross section at the estimation of the fusion probability $P_{\rm CN}$ from the analysis
   of the  experimental data.
  \begin{equation}
  P_{\rm CN}=\frac{\sigma_{ER}+\sigma_{fusion-fission}}{\sigma_{ER}+\sigma_{fusion-fission}+\tilde{\sigma}_{\rm qf}},
  \label{PCN}
  \end{equation}
  where $\sigma_{ER}$ and $\sigma_{fusion-fission}$ are the evaporation residue and
  fusion-fission cross sections, respectively; the cross section of quasifission estimated
  by the assumption of experimentalists is
  $\tilde{\sigma}_{\rm qf}=\sigma_{asymqf}+\sigma_{symqf}$. For example, the experimental data
  of capture cross section presented in Ref. \cite{Knyazheva} were compared with the theoretical
  results in Ref. \cite{ANPRC09}. The  separation of
   the mass symmetric products  of quasifission  $\sigma_{symqf}$ from the
   fusion-fission products is the other interesting
   and difficult problem of future theoretical and experimental studies
   \cite{GGJPCS282,ANJPCS282,MTEPJA53}.

    Therefore, the experimental value of $P_{\rm CN}$ is larger than its theoretical value.
  This means that the estimated capture cross section from the analysis of
  experimental data is smaller than theoretical capture cross section
  calculated as a sum of the full momentum transfer events.
Theoretical values of the capture cross sections are calculated with the quantities
characterizing the entrance channel  by formula
\begin{equation}
\sigma_{cap}(E_{\rm c.m.})=\frac{\lambdabar^2}{4\pi}\sum\limits^{\ell_d}_{\ell=0} (2\ell+1)
\mathcal{P}^{(\ell)}_{cap}(E_{\rm c.m.}),
\label{defcap}
\end{equation}
where $\mathcal{P}^{(\ell)}_{cap}(E_{\rm c.m.})$ is the capture probability of the projec-tile-nucleus
by the target-nucleus in collision with energy  $E_{\rm c.m.}$ and orbital angular momentum
$L=\hbar \ell$; $\mu$ is the reduced mass of colliding nuclei and $\lambdabar=\hbar/\sqrt{2\mu E_{\rm c.m.}}$.
All partial waves corresponding to the full momentum transfer events are included into the
 summation in Eq.(\ref{defcap}). This means that
Eq.(\ref{defcap}) includes the yield of projectile- and target-like  products  together with
fusion-fission, quasifission and evaporation residue products.
The DNS formed in the collisions with the given values of $E_{\rm c.m.}$
and $\ell$ evolves to complete fusion due to the transfer of all nucleons of the light fragment to the
heavy one or it can decay forming binary products with charge and mass numbers in the wide
 range.  According to our view, the  projectile- and target-like  products having low total
  kinetic energy are considered as the quasifission products. The dynamical calculation by the
   method presented in Section \ref{model} allows us  to find angular momentum distribution of
  the DNS formed in capture.
 In some methods of capture calculations, the variation of the maximum value of the orbital angular momentum
 $\ell_d$ or another way is  used to reach  an agreement with the experimental values of the capture cross
  section which is found by ignoring the yield of the capture products which have close
values to the initial mass and charge numbers of colliding nuclei \cite{AtDat17}.

\section{Theoretical model}
\label{model}

\begin{figure}
\vspace*{0.0cm}
\hspace*{-0.4cm}
\resizebox{0.65\textwidth}{!}{\includegraphics{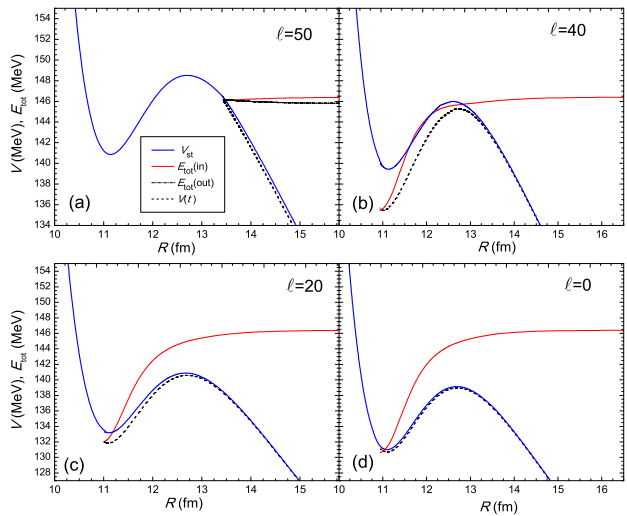}}
\vspace*{-2.80cm} \caption{(Color online) Deep inelastic collision (a)
and capture (b),(c),(d) trajectories at collision energy $E_{\rm
c.m.}=146.41$ MeV for the $^{36}$S+$^{206}$Pb reaction. The thick solid
lines are the interaction potential $V_(t)$; thin solid  and dot-dashed
 lines are the total energy $E_{\rm tot}$ of the relative motion for the incoming and outgoing
trajectories (only for the case of $L=50 \hbar$), respectively; the dashed lines are the time dependent
interaction potential $V(t)$ calculated taking into account damping
angular momentum and nucleon exchange between nuclei \cite{NFTA05}.
The total energy $E_{\rm tot}$ decreases  due to dissipation by the radial friction
 coefficient and the
 dynamical interaction potential $V(t)$ presented by the dotted line
 while the nucleus-nucleus potential including the DNS rotational energy
 calculated with the undamped values of $L$ is shown by the thick solid line.
} \vspace*{10pt}
\label{deepcap}
\end{figure}

The dynamics of complete fusion in heavy ion collisions at low energies is determined
by the orbital angular momentum, the charge and mass numbers, shape, orientation angles
 of their symmetry axis  and the shell
structure of the interacting nuclei.
 The probability of the mass and charge distributions between fragments of the DNS
 and the probability of its decay depend on the shell structure, the excitation energy
 and angular momentum of the system. Therefore, it is important to include
 into consideration the construction of the theoretical methods to study
 the role of the entrance channel in formation of the reaction products.
 The examples of results corresponding to the full
 momentum transfer events are presented  in panels (b),(c) and (d) of Fig. \ref{deepcap}.

\subsection{Calculation of capture cross section}

\begin{figure}
\vspace*{0.0cm}
\begin{center}
\resizebox{0.51\textwidth}{!}{\includegraphics{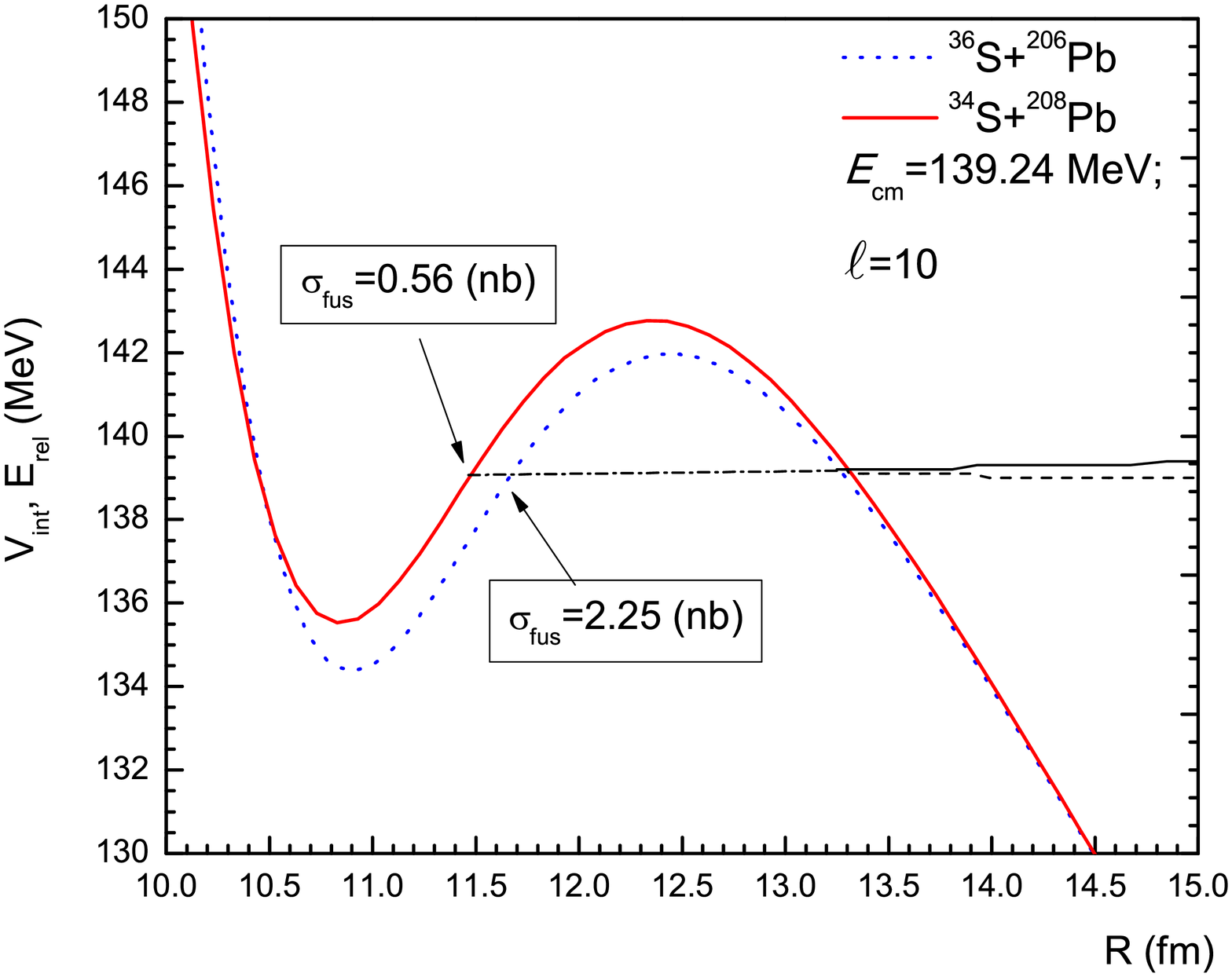}}
\end{center}
\vspace*{-1.15cm} \caption{(Color online) Deep-inelastic collision
(dashed line),
sub-barrier capture (dot-dashed line) and nucleus-nucleus potential for
$^{36}$S+$^{206}$Pb (dotted line) and $^{34}$S+$^{208}$Pb (solid
line).} \label{subfus}
\end{figure}

Two conditions must be satisfied for the capture: 1) the initial energy
$E_{\rm c.m.}$  of a projectile in the center-of-mass system should be
enough to reach the potential well of the nucleus-nucleus interaction
 (Coulomb barrier + rotational energy of the entrance channel) by
 overcoming or tunneling through the barrier along relative distance
 in the entrance channel; 2) at the same time the value of the relative
 kinetic energy above the entrance channel barrier should in correspondence
 with the size of the potential well: in case of the collision of the
 massive nuclei the size of the potential is small and, if the initial
 collision energy is very large relative to the the entrance channel barrier,
  the dissipation of the kinetic energy may be not enough to make
  its value lower than barrier of potential well, {\it i.e.} to
  cause trapping into potential well. As a result, the capture does not occur
 and the deep-inelastic collision takes place
  (as in Fig. 2b of Ref. \cite{NFTA05} or Fig. 1 of Ref. \cite{GHMN00}).
If there is no potential well, the
deep-inelastic collision takes place only. In this work the possibility
of capture by tunnelling through the Coulomb barrier, at the collision energies
($E_{\rm c.m.}$) lower than the barrier, is taken into account. In this case, at the same
values of the angular momentum both capture and deep-inelastic collision
can occur.
Fig.~\ref{deepcap} shows the dependence of the total energy ($V(t)+E_{\rm kin}$)
of the radial motion $E_{\rm tot}$ and nucleus-nucleus potential on the distance $R$
between centers of nuclei for the $^{36}$S+$^{206}$Pb reaction
 at the collision energy $E_{c.m.}=146.41$ MeV.
 We should note that the capture probability is calculated for all values of $\ell$
 including the case presented in Fig. \ref{deepcap}. The capture probability
 by tunneling in the case showed in Fig. \ref{deepcap}(a) is discussed later
 in Fig. \ref{subfus}.
 The nucleus-nucleus potential $V(\ell,\{\alpha_i\};R)$ consists of three
parts as
\begin{eqnarray}
V(\ell,\alpha_1,\alpha_2; R) &=& V_{\rm Coul}(\alpha_1,\alpha_2; R) \\
&+& V_{\rm nucl}(\alpha_1,\alpha_2; R)
 + V_{\rm rot}(\ell,\alpha_1,\alpha_2; R),\nonumber
\end{eqnarray}
where $V_{\rm Coul}$, $V_{\rm nucl}$ and $V_{\rm rot}$ are the
Coulomb, nuclear and rotational potentials, respectively.
In the case of collision of nuclei with the deformed shape in their ground states,
the dependence of the nucleus-nucleus potential on orientation angle of their axial
symmetry axis should be taken into account.  We refer to
Ref.~\cite{FGLM04} and Appendix~A of Ref.~\cite{NFTA05} for the
detailed expressions of these potentials in terms of the orientation
angles of the symmetry axis of the colliding nuclei.

The colliding nuclei in the $^{34}$S+$^{208}$Pb and $^{36}$S+$^{206}$Pb reactions
are spherical in their ground states, therefore, possibility of the population
of vibrational states at their excitation is considered.
 As the amplitudes of the surface vibration we use deformation parameters of first excited
 2$^+$ and $3^-$ states of the colliding nuclei. The values of the deformation parameters of
  first  excited 2$^+$ and $3^-$ states  are presented in Table \ref{tabdeform} which are taken
   from Ref(s). \cite{Raman} ($\beta^+_2$) and \cite{Spear} ($\beta^-_3$).

\begin{table}[h]
\caption{Deformation parameters $\beta$ and mean lifetime  $\tau$ of first excited
 2$^+$ and $3^-$ states of the colliding nuclei used in the calculations in this work.
 \label{tabdeform}}
\begin{center}
\begin{tabular}{|c|c|c|c|c|}
 \hline
 Nucleus      &    $^{34}$S &  $^{36}$S & $^{206}$Pb & $^{208}$Pb \\
  \hline
  $\beta^+_2$ \cite{Raman}  & 0.252           & 0.168     & 0.0322 & 0.055 \\
  $\tau_{2^+}(10^{-12}$s) \cite{Raman} & 0.023 & 0.110     & 0.11   & 0.0012 \\

  $\beta^-_3$ \cite{Spear}     &      0.330            &   -     & 0.083  & 0.100  \\
  $\tau_{3^-}(10^{-12}$s) \cite{Spear} &   $0.130$     & - &      -       & $47$ \\

  \hline
  \end{tabular}
  \end{center}
  \end{table}

The surface vibrations are regarded as independent harmonic vibrations and the nuclear radius
is considered to be distributed as a Gaussian distribution~\cite{EsbensenNPA352},
\begin{equation}
g(\beta_2, \beta_3; \alpha) = \exp
\left[ -\frac{(\sum_{\lambda}\beta_{\lambda} Y_{\lambda 0}^* (\alpha))^2}{2 \sigma_{\beta}^2} \right] (2\pi \sigma_{\beta}^2)^{-1/2},
\end{equation}
where $\alpha$ is the direction of the spherical nucleus. For simplicity, we use $\alpha=0$:

\begin{equation}
\sigma^2_{\beta} = R_0^2 \sum_{\lambda}\frac{2\lambda + 1}{4\pi} \frac{\hbar}{2D_\lambda \omega_\lambda} = \frac{R_0^2}{4\pi} \sum_{\lambda} \beta_\lambda^2,
\end{equation}
where $\omega_{\lambda}$ is the frequency and $D_{\lambda}$ is the mass parameter of a collective mode.

The first step at the estimation of the capture cross section is
the calculation of the partial capture cross sections for the
seven values for each deformation parameters $\beta_2$ and $\beta_3$
 in the corresponding ranges $-\beta^{+}_2 <\beta_2< \beta^{+}_2$ and
 $-\beta^{-}_3 <\beta_3< \beta^{-}_3$ for the vibrational nuclei, {\it i.e.}
 the differences between intermediate values of the deformation parameters
 used in calculations are $\Delta\beta^{+}_{2}=\beta^{+}_{2}/3.$ and  $\Delta\beta^{+}_{2}=\beta^{+}_{2}/3.$,
 respectively.
This procedure is acceptable since the mean lifetime  $\tau$ of first excited
 2$^+$ and $3^-$ states (see Table 1) are much larger than interaction time of colliding
 nuclei at capture and complete fusion times which do not precede $10^{-19}$ s.
 Therefore, deformation parameters $\beta_{2}$ and $\beta_{3}$ can be considered
  as the frozen values during the capture process.

The partial capture cross-section
$\sigma^{(\ell)}_{cap}(E_{\rm c.m}, \{\beta_i\})$
is determined by calculation of the capture probability \\
$\mathcal{P}^{(\ell)}_{cap}(E_{\rm c.m.},\{\beta_i\})$
of trapping the curve presenting the dependence of total kinetic energy
on the time dependent internuclear distance into the potential well of
the nucleus-nucleus interaction:
\begin{equation}
\label{sigmacap}
\sigma^{(\ell)}_{cap}(E_{\rm
c.m.},\{\beta_i\})=\frac{\lambda^2}{4\pi}(2\ell+1)\mathcal{P}^{(\ell)}_{cap}(E_{\rm c.m.},\{\beta_i\}).
\end{equation}
Here $\lambda$ is the de Broglie wavelength of the entrance channel.
The capture probability  $\mathcal{P}^{(\ell)}_{cap}(E_{\rm
c.m.},\{\beta_i\})$, which is calculated by classical equation of motion,
 is equal to $1$ or $0$ for given beam energy
and orbital angular momentum. In dependence on the beam energy,
$E_{\rm c.m.}$, there is a $\ell$-window ($ \ell_m < \ell < \ell_d$)
for capture as a function of orbital angular momentum:
\begin{equation}
\mathcal{P}^{(\ell)}_{cap}(E_{\rm c.m.},\{\beta_i\}) = \left\{
\begin{array}{lll}
1, \mbox{if } \ell_m < \ell < \ell_d \mbox{ and } \nonumber\\
\hspace*{1.5cm} E_{\rm c.m.} > V_{B}\, , \\
0, \mbox{if }\ell < \ell_m \mbox{ or }\ell>\ell_d \mbox{ and }
\nonumber\\
\hspace*{1.5cm} E_{\rm c.m.}>V_{B}\, , \\
\mathcal{P}^{(\ell)}_{WKB}, \mbox{for all } \ell \mbox{ if } E_{\rm
c.m.}\leq V_{B}\, , \\
  \end{array} \right.
  \label{CapClass}
\end{equation}
where $\ell_m$ and $\ell_d$ are the minimum and maximum values of the
orbital angular momentum $\ell$ leading to capture at the given collision
energy; $V_B$ is the barrier of the nucleus-nucleus potential in the entrance channel;
$\mathcal{P}^{(\ell)}_{WKB}$ is probability of the barrier penetrability
which is calculated by the formula is derived from the WKB approximation
(see Eq. \ref{tunnel}).
The absence of capture at
$\ell < \ell_m $ means that the total energy
curve as a function of $E_{\rm c.m.}$ is not trapped into potential well:
dissipation of the initial kinetic energy is not enough to be
trapped due to the restricted value of the radial friction coefficient.
The number of partial waves giving a contribution to the capture is
calculated by the solution of Eq(s) (\ref{maineq1})-(\ref{maineq5}) for the radial and
orbital motions  simultaneously.

 The collision trajectory, rotational angle,
angular velocity and the moment of inertia of the DNS formed after
capture for a given beam energy $E_{\rm c.m.}$ and orbital angular
momentum $L_0$ are found by solving the following equations of
motion~\cite{NFTA05,AJNM96}:
\begin{eqnarray}
 \label{maineq1}
 \mu(R)\frac{d \dot R}{dt} &+&
 \gamma_{R}(R)\dot R(t)= F(R),\\
 \label{maineq2}
 F(R)&=&
 -\frac {\partial V(R)}{\partial R}-
 \dot R^2 \frac {\partial \mu(R)}{\partial R}\,,\\
 \label{maineq3}
 \frac{dL}{dt}&=&\gamma_{\theta}(R)R(t)\left(\dot{\theta}
 R(t) -\dot{\theta_1} R_{\rm 1eff} -\dot{\theta_2} R_{\rm 2eff}\right), \\
 \label{maineq4}
 \frac{dL_1}{dt}&=&\gamma_{\theta}(R)\left[R_{1eff}\left(\dot{\theta}
 R(t) -\dot{\theta_1} R_{\rm 1eff} -\dot{\theta_2} R_{\rm 2eff}\right) \right.\nonumber\\
  &-& \left.2 a (R_{\rm 1eff}\dot{\theta_1}-R_{\rm 2eff}\dot{\theta_2})\right], \\
 \label{maineq5}
 \frac{dL_2}{dt}&=&\gamma_{\theta}(R)\left[R_{\rm 1eff}\left(\dot{\theta}
 R(t) -\dot{\theta_1} R_{\rm 1eff} -\dot{\theta_2} R_{\rm 2eff}\right)\right. \nonumber\\
   &+&\left. 2 a(R_{\rm 1eff}\dot{\theta_1}-R_{2eff}\dot{\theta_2})\right], \\
 L_0&=&L(\dot{\theta})+L_1(\dot{\theta_1})+L_2(\dot{\theta_2})\,,\\
 L(\dot{\theta})&=&J_{\rm DNS}(R) \dot{\theta};\,\\
L_1(\dot{\theta_1})&=&J_1 \dot{\theta_1};\,\\
L_2(\dot{\theta_2})&=&J_2 \dot{\theta_2},\\
 E_{\rm rot}&=&\frac{J_R(R) \dot{\theta_{}}{}^2}2+\frac{J_1
 \dot{\theta_1}^2}2+\frac{J_2 \dot{\theta_2}^2}2\,,
 \end{eqnarray}
where    $J_{\rm DNS}$ is the DNS moment of inertia and it is calculated
by the rigid-body approximation as
\begin{eqnarray}
\label{DNSJ} J_{\rm DNS}(\beta_{i_1}^{},\beta_{i_2}^{};R) = \mu(R) \,
R^2(\beta_{i_1}^{},\beta_{i_2}^{}) + J_1 + J_2,
\end{eqnarray}
where $R(\beta_{i_1}^{},\beta_{i_2}^{})$ is the distance between the
centers of nuclei at their given vibrational states $i_1$
and $i_2$ of colliding nuclei; $J_R$ and
$\dot\theta$, $J_1$ and $\dot\theta_1$, $J_2$ and $\dot\theta_2$ are
the moments of inertia and angular velocities of the DNS and its
fragments, respectively. We also defined $R_{\rm 1eff} = R_1 + a$
and $R_{\rm 2,eff} = R_2 + a$, where $R_1$ and $R_2$ are the radius
of the interacting nuclei with $a = 0.54$~fm~\cite{NFTA05}. Here, $L_0$
and $E_{\rm rot}$ are determined by the initial conditions.
The initial value of the relative distance $R$ is taken
equal to 20 fm; the initial values of the orbital angular momentum $L_0$
is given in the range $0\div80 \hbar$ by the step $\Delta L=5 \hbar$
since fission barrier disappears at $\Delta L=60 \hbar$; the initial
velocity of the projectile is determined by the values of $E_{\rm c.m.}$
and $L_0$.

 The nucleus-nucleus interaction potential, radial and tangential
friction coefficients and inertia coefficients are calculated in the
framework of our model \cite{NFTA05,GHMN00,FGLM04,NKO16}.

It should be stressed that friction coefficients $\gamma_R$ and
 $\gamma_{\theta}$ of the relative motions are sensitive to the
 shape of the colliding nuclei (see Ref.~\cite{NFTA05}).
 For simplicity in the presentation of formulas the following labels
 are used in Eq(s) (9)-(18) of motions
  $F(R)=F(R,\{\beta_i\})$, $\gamma_R(R)=\gamma_R(R,\{\beta_i\})$,
  $\gamma_{\theta}(R)=\gamma_{\theta}(R,\{\beta_i\})$,
  $V(R)=V(R,\{\beta_i\})$,  where $R(t)$ is the relative distance,
   $\dot R(t) \equiv dR(r)/dt$ is the corresponding velocity.

In sub-barrier capture processes, the
barrier penetrability formula is derived from the WKB approximation and
it is calculated by:
\begin{equation}\label{penetration}
\mathcal{P}^{(\ell)}_{WKB}(E_{\rm c.m.},\{\beta_i\})=
\exp\left[-2\int_{R_{in}}^{R_{out}}k(R,\ell,\{\beta_i\})dR\right],
\label{tunnel}
\end{equation}
where
 \begin{equation}
k(R,\ell,\{\beta_i\})=\sqrt{\frac{2\mu}{\hbar^2}(V(R,\ell,\{\beta_i\})-E_{\rm c.m.})}.
\end{equation}

 $R_{in}$ and $R_{out}$ are inner and outer turning points which were estimated by
$V(R)=E_{\rm c.m.}$.

The second stage is an averaging by the expression (\ref{gauss})
   to find an averaged value of the partial capture cross section over
   surface  vibrational state:
\begin{eqnarray}
\langle \sigma^{(\ell)}_{\rm cap} (E_{c.m}) \rangle &=& \int^{\beta_{2+}}_{-\beta_{2+}}
\int^{\beta_{3-}}_{-\beta_{3-}} \sigma^{(\ell)}_{\rm cap}(E_{\rm c.m}, \beta_2, \beta_3) \nonumber\\
&\times & g(\beta_2, \beta_3) d\beta_2 d\beta_3.
\label{gauss}
\end{eqnarray}
 The deformation parameters of the vibrational states can be considered as frozen during the capture process
 since as it is seen from the Table 1 that the mean lifetime of the first excited states $2^+$ and
 $3^-$  is much longer than the time scale of capture and fusion processes. The time scale of the
  capture and fusion processes is less than 10$^{-19}$ s.

 Comparison of the capture probabilities calculated for the $^{36}$S+$^{206}$Pb
 (dotted line)
 and $^{34}$S+$^{208}$Pb (solid line) reactions at the collision energy
 $E_{\rm c.m.}$=139.24 MeV and $\ell=$10 is presented in Fig. \ref{subfus}.
 The capture probability for the former reaction is about 4 times larger
 than the one of the latter reaction since the potential barrier calculated
 for the latter reaction is higher and thicker than the one for the
 former reaction since the projectile
  $^{36}$S has larger $N/Z$ - ratio  than  $^{34}$S.

\begin{figure}
\vspace*{-1.0cm}
\hspace*{-0.15cm}
\begin{center}
\resizebox{0.62\textwidth}{!}{\includegraphics{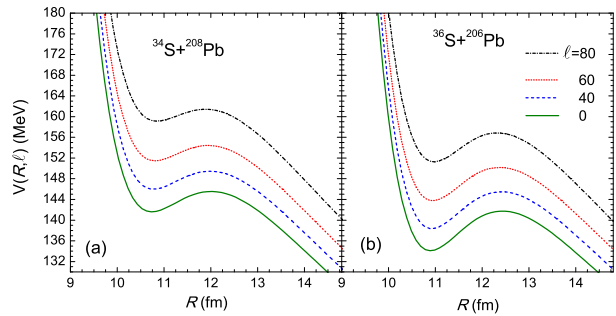}}
\end{center}
\vspace*{-4.5cm}
\caption{(Color online) The influences of the rotational angular
momentum on nucleus-nucleus potential; solid line for $\ell=0$;
 dashed line for $\ell=40$; dotted line for $\ell=60$;
 dash-dotted line for $\ell=80$ at deformation parameters
 presented in Table 1.} \label{potenell}
\end{figure}

The role of the initial orbital angular momentum $\ell$ in the
heavy ion collisions can be seen from the Fig.\ref{potenell}. This
figure represents the dependence of the depth of the potential well and
 the Coulomb barrier as functions of the the orbital angular momentum
for $^{34}$S+$^{208}$Pb and $^{36}$S+$^{206}$Pb reactions. It can be
clearly seen that the increase of $\ell$  can lead to reduction
of the potential well.

  The value of the entrance channel barrier (Coulomb barrier at $\ell$=0)
  calculated for the $^{34}$S+$^{208}$Pb reaction is higher than the
  one obtained for the $^{36}$S+$^{206}$Pb reaction. The low barrier
  of the entrance channel is favorable to decrease the CN excitation
  energy since it makes lower threshold value of the beam energy
  leading to the CN formation.

The total capture cross section is found by summarizing over
partial waves:
\begin{eqnarray}\label{sumcap}
\sigma_{cap}(E_{\rm c.m.})=\sum\limits^{\ell=\ell_d}_{\ell=\ell_m}\langle\sigma^{(\ell)}_{cap}(E_{\rm c.m.})\rangle.
\end{eqnarray}
It should be noted the range of orbital angular momentum values $\ell_m<\ell<\ell_d$
 contributing to
capture cross section depends on the collision energy  $E_{\rm c.m.}$.
The calculations have shown that the minimum value of angular
momentum is zero ($\ell_m$=0) for the  $^{34}$S+$^{208}$Pb
and $^{36}$S+$^{206}$Pb reactions.

\subsection{Evolution of the DNS}
\label{results2}

The evolution of DNS to complete fusion or quasifission is determined by the landscape of
potential energy surface (PES) and nucleon distribution in the single-particle states
of the interacting nuclei.
The intrinsic fusion $B^*_{\rm fus}(Z,A)$ and quasifission $B_{qf}(Z,A)$ barriers,
 as well as the excitation energy $E^*_{Z,A}$ of DNS are found from PES. In calculation of
 the fusion probability, these  $B^*_{\rm fus}(Z,A)$ and $B_{qf}(Z,A)$ barriers are main
  quantities  together with the DNS excitation energy $E^*_{\rm DNS}(Z,A)$ being functions of
   the mass and  charge asymmetry of the DNS configuration.

The values of $U_{dr}(Z,R_m)$
as a function of angular momentum $\ell$ are found from the data of
PES calculated by the formula
and the reaction energy balance ($Q_{gg}$)
corresponding to the charge asymmetry configuration $Z$ of DNS.
\begin{equation}
U_{dr}(Z,A,\ell,R_m)=Q_{gg}+V(Z,A,\ell,R_m)
\end{equation}
where $Z=Z_1$ and $A=A_1$ are charge and mass numbers of a DNS
fragment while the ones of another fragment are $Z_2=Z_{\rm
tot}-Z_1$ and $A_2=A_{\rm tot}-A_1$, where $Z_{\rm tot}$  and
$A_{\rm tot}$ are the total charge and mass numbers of a reaction,
 respectively; $Q_{\rm gg}$ is the reaction energy balance
used to determine the excitation energy of CN: $Q_{\rm
gg}=B_1+B_2-B_{\rm CN}$. The binding energies for the initial projectile
and target nuclei ($B_1$ and $B_2$) are obtained from the mass
tables in Ref. \cite{MassAW95}, while the one of CN  ($B_{\rm CN}$)
are obtained from the mass tables \cite{MolNix,Muntian03}. If there
is no potential well of $V(Z,A,R,\ell)$ at large values of angular
momentum or for symmetric massive nuclei, we use  $R_m$
corresponding to the smallest value of the derivation $\partial
V(Z,A,R_m,\ell)/\partial R$ in the contact area of nuclei.
The intrinsic fusion barrier,
$B^{\ast}_{fus}(Z,A,\ell)$, is determined as the difference between
the maximum value of the driving potential between $Z = 0$ and $Z =
Z_P$ and the initial charge value ,
\begin{equation}\label{Bfus}
B^{\ast}_{fus}(Z,A,\ell)=U^{max}_{dr}(\ell)-U_{dr}(Z_P,A_P,\ell)
\end{equation}
where $U^{max}_{dr}(\ell)=U_{dr}(Z_{max},A_{max},\ell)$.

These main  quantities
   are found from the analysis of PES and by the use of collision energy $E_{\rm c.m.}$
   (see \cite{NFTA05,KKNMG15,FGRMN05}).
   Due to nucleon exchange between DNS nuclei
 their mass and charge distributions are changed as functions of time.
 Their evolution are estimated by solving the transport master equation with
 the transition coefficients calculated microscopically \cite{Anton94,DiasPRC64}.
 The proton and neutron systems of nuclei have own energy scheme of the single-particle
states and the single-particle schemes depend on the mass and charge numbers
of nuclei. Consequently, the transition coefficients
$\Delta^{(-)}_{K}$ and $\Delta^{(+)}_{K}$ of the transport master equation (\ref{massdec})
being sensitive to the energy scheme and occupation numbers of the single-particle states
(see Eq. (\ref{delt})) of the interacting nuclei depend on the mass numbers too.
 The dependence of the transition coefficients $\Delta^{(-)}_{K}$ and $\Delta^{(+)}_{K}$
 on the mass and charge numbers of nuclei leads to the correlation between proton and neutron
  numbers in them.

The difference between the mass and charge distributions at the given time of the DNS
evolution depends on the initial $N/Z$ - ratio in colliding nuclei
 since transition coefficients causing nucleon transfer are different for the isotopes
  with different neutron numbers of the nucleus with the same charge numbers.
  The difference in the mass and charge evolutions for  the $^{34}$S+$^{208}$Pb and
  $^{36}$S+$^{206}$Pb reactions leads to the difference in fusion probabilities
  in these reactions.
The difference in fusion cross section immediately causes difference
in the evaporation residue cross sections measured
in these reactions \cite{KNHAB12}.

The mass and charge distributions among the DNS fragments are
calculated by solving the transport
master equation:
\begin{eqnarray}
\label{massdec} \frac{\partial}{\partial t} P_{K}^{}(E^*_K,\ell,t)
&=& \Delta^{(-)}_{K+1} P_{K+1}^{} (E^*_{K+1},\ell,t)\\
&+& \Delta^{(+)}_{K-1}
P_{K-1}^{} (E^*_{K-1},\ell,t) \nonumber\\  \mbox{} &-&
\left(\Delta^{(-)}_{K}+\Delta^{(+)}_{K}+\Lambda^{\rm qf}_{K} \right)
P_{K}^{}(E^*_K,\ell,t)\nonumber
\end{eqnarray}
for $K=Z,N$ (for proton and neutron transfers). Here $A_1=A=N+Z$ is the mass number
of the light fragment of DNS while $A_2=A_{CN}-A$ and $Z_2=Z_{CN}-Z$ are
the mass and charge numbers of the heavy fragment of DNS;
$P_K^{}(A,E^*_{DNS}(t),\ell,t)$ is the probability of population of
the configuration $(K, K_{\rm CN}-K)$ of the DNS at the given values of $E^*_{DNS}(t)$,
 $\ell$ and interaction time $t$.
 To make easy writing of the Eq(s).(\ref{massdec})  we have used
 the following designations:
 \begin{eqnarray}
&&P_{K}^{}(E^*_K,\ell,t)= P_{K}^{}(A,E^*_K,\ell,t),\nonumber\\
&&P_{K\pm1}^{}(E^*_K,\ell,t)= P_{K\pm1}^{}(A\pm1,E^*_K,\ell,t),\nonumber\\
&&\Delta^{(\mp)}_{K}=\Delta^{(\mp)}_{K}(A),\nonumber\\
&&\Delta^{(\mp)}_{K\pm1}=\Delta^{(\mp)}_{K\pm1}(A\pm1),\nonumber\\
&&\Lambda^{\rm qf}_{K}=\Lambda^{\rm qf}_{K}(A)\nonumber\\
&&E^*_K=E^*(K,A,\ell)\nonumber
 \end{eqnarray}
Note these quantities and all quantities characterizing the single-particle states
 $\tilde\varepsilon$, $n^{(K)}_{P,T}$ and matrix elements $g^{(K)}_{PT}$
 in Eq.(\ref{delt}) depend on the mass numbers $A$ and $A_2=A_{CN}-A$ of the light and heavy fragments, respectively.
The transition coefficients of multinucleon transfer  are calculated as in \cite{JMN86}
\begin{eqnarray}
\label{delt} \Delta^{(\pm)}_{K}(A) &=& \frac{4}{\Delta t}
\sum\limits_{i_P,j_T}|g^{(K)}_{i_Pj_T}(A)|^2 \nonumber\\
&\times& n^{(K)}_{j_T,i_P}(A,t) \ \left( 1 -
n^{(K)}_{i_P,j_T}(A,t) \right) \nonumber \\
&\times& \mbox{}
 \frac{\sin^2 [ \Delta t(\tilde\varepsilon_{i_P}^{(K)}(A)-
 \tilde\varepsilon_{j_T}^{(K)}(A))/2\hbar ]}
 {(\tilde\varepsilon_{i_P}^{(K)}(A) - \tilde\varepsilon_{j_T}^{(K)}(A))^2},
\end{eqnarray}
where the matrix elements $g^{(K)}_{i_Pj_T}(A)^{}$  describe one-nucleon exchange
between the DNS nuclei ``$P$'' and ``$T$'' (for the proton exchange $K=Z$
and for the neutron exchange  $K=N$) and their values are calculated
microscopically as in Ref.\cite{AJN92}. Due to dependence of the transition coefficients
$\Delta^{(-)}_{K}$ and $\Delta^{(+)}_{K}$ on the mass and charge numbers
of nuclei the neutron and proton distributions $P_Z$ and $P_N$ are
correlated since their master equations are solved parallel way but
consequently with the time step $\Delta t$.
 It is clear that the proton and neutron transfers
 takes place simultaneously but with the different probabilities.
 The letters ``$P$'' and ``$T$'' are used to indicate
the single-particle states of nucleons in projectile-like (light) and
target-like (heavy) fragments, respectively, of DNS.
 In the present work, we follow the scheme of
Ref.~\cite{AJN92} for estimating these values with $\Delta
t=10^{-22} \mbox{ s} \ll \,t_{\rm DNS}$, where $t_{\rm DNS}$ is the
interaction time of the DNS nuclei and according to
calculations it has values $t_{\rm DNS} > 5\cdot 10^{-22}$ s.
 This way allows us to take into account non-equilibrium distribution
 of the excitation energy between the fragments by
 in calculation of the single-particle occupation numbers $n_{i_P}^{(K)}$
and $n_{i_T}^{(K)}$ following Ref.~\cite{AJN94}.
The excitation of the DNS is calculated  by the estimation of the population of the
 proton and neutron hole states of one fragment under influence of the mean-field
 of the other fragment.
 This kind of evolution of the single-particle
 occupation numbers $n_{i_P}^{(K)}$ and $n_{i_T}^{(K)}$ is established
 by solution of the  Liouville quantum equation for the occupation numbers
 with the linearised collision term:
 \begin{equation}
 i \hbar \frac{\partial \tilde n_{i_P}^{(K)}(t)}{\partial t}=[H,\tilde n_{i_P}^{(K)}]+
 \frac{\tilde n_{i_P}^{(K)}(t)- n_{i_P}^{\rm eq(K)}(T_Z)}{\tau_{i_P}^{(K)}},
 \label{Liouvil}
 \end{equation}
 where $H$ is the sum of the collective Hamiltonian $H_{\rm rel}$ of the relative motion
 of interacting nuclei of DNS,
 the secondary quantized Hamiltonian $H_{\rm in}$ of the intrinsic motion of nucleons
 in them  and the coupling term $V_{\rm int}$ corresponding to the interaction between
 collective relative motion of nuclei and intrinsic motion of nucleons,
\begin{equation}
H=H_{\rm rel}+H_{\rm in}+V_{\rm int}.
\end{equation}
  The last term $V_{\rm int}$ is responsible to excitation  of the DNS fragments and
  it leads to evolution of the occupation numbers of nucleons.
     The use of the linearised collision term
 in Eq. (\ref{Liouvil}) allows us to determine the time dependent occupation numbers
 evolve to  the thermal equilibrium ones $n_{i_P}^{\rm eq(K)}(T_Z)$; $\tau_{i_P}^{(K)}$
  is the relaxation time
 of the  excited  single-particle state $i_P$  of the light fragment ``$P$'' ($i_T$ for heavy fragment ``$T$'').
  The details of calculation can be find in Refs.\cite{AJNM96,AJN94}.
 The thermal equilibrium occupation numbers are calculated by the usual expression:
 \begin{equation}
 n^{\rm eq}(T_Z)=\frac{1}{1+\exp[\frac{(\tilde\varepsilon_{P_K}-\varepsilon_{F})}{T_Z}]},
 \end{equation}
 where $T_Z$ is the effective temperature of DNS with the charge asymmetry $Z$ and its value
  is determined by the excitation energy $E*_Z$ of DNS as the Fermi-gas temperature
   $T=\sqrt{\frac{E^*_Z}{a}}$
where $a=1/12$ MeV$^{-1}$.  $E^*_Z$ is the excitation energy
 of DNS and it is determined by the initial beam energy and the minimum
 of the potential energy as
\begin{equation}
\label{Edns} E^*_Z(A,\ell) = E_{\rm c.m.}-V(Z,A,R_m,(\ell))+\Delta
Q_{\rm gg}(Z,A),
\end{equation}
where  $V(Z,A,R_m,(\ell))$ is the minimum value of the potential well
$V(Z,A,R,\ell)$ at $R_m$;  $\Delta Q_{\rm gg}(Z,A)=B_1+B_2-B_P-B_T$
is included to take
into account the change of the intrinsic energy of DNS due to
nucleon transitions during its evolution along mass and charge
asymmetry axes, where $B_1$, $B_2$, $B_P$ and $B_T$ are binding energies of the
initial (``1'' and ``2'') and interacting fragments (``P'' and ``T'') at the
given time $t$ of interaction.

$\tilde\varepsilon_{P_K}^{}$ and $\tilde\varepsilon_{T_K}^{}$ are
perturbed energies of single-particle states:
$\tilde\varepsilon_i^{}=\varepsilon_i^{}+V_{ii}$, $V_{ii}$ is the
diagonal elements of the matrix $V_{ii'}$ (see details in Ref(s).
~\cite{Anton94,AJN94}). In Eq.~(\ref{yield}), $\Lambda^{\rm qf}_{Z}$
is the Kramer's rate for the decay probability of the DNS
 into two fragments with charge numbers $Z$ and $Z_{\rm
CN}-Z$~\cite{AAS03,EPJA34}, which is proportional to $\exp [-B_{\rm
qf}(Z)/(kT)]$.
The decay probability increases by decreasing
the quasifission barrier, $B_{\rm qf}$, which is taken equal to
the depth of the potential well $V(Z,A,R,\ell)$ presented in Fig(s).
  \ref{deepcap}, \ref{subfus} and \ref{potenell}.

\begin{figure}
\vspace*{-1.5cm}
\hspace*{-0.15cm}
\begin{center}
\resizebox{0.70\textwidth}{!}{\includegraphics{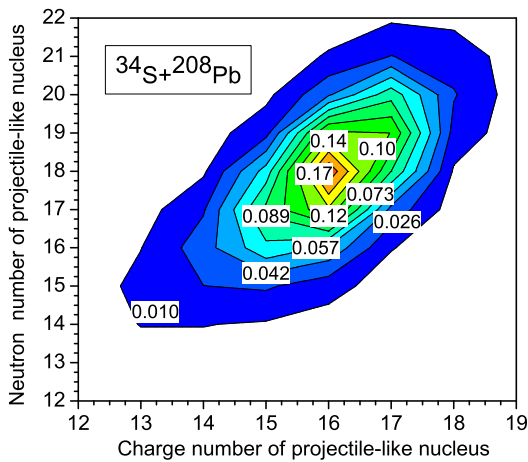}}
\end{center}
\vspace*{-2.5cm}
\caption{(Color online) The dependence of the neutron distribution
as a function of the charge number of the light fragment of the
dinuclear system formed in the $^{34}$S+$^{208}$Pb reaction.}
\label{NZ34S208Pb}
\end{figure}

\begin{figure}
\vspace*{-1.5cm}
\hspace*{-0.15cm}
\begin{center}
\resizebox{0.70\textwidth}{!}{\includegraphics{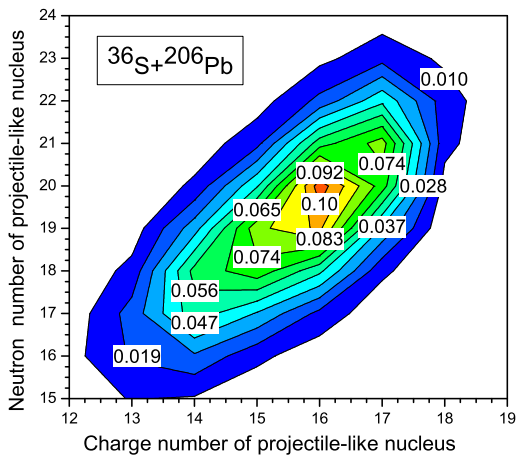}}
\end{center}
\vspace*{-2.5cm}
\caption{(Color online) The same dependence of the neutron distribution
as a function of the charge number of the light fragment of the
dinuclear system formed in the $^{36}$S+$^{206}$Pb reaction.}
\label{NZ36S206Pb}
\end{figure}

 It should be noted that
 the fusion probability $P_{\rm CN}$ is a function of the mass and charge asymmetry
 of the DNS nuclei and therefore, the contributions to the complete
 fusion of different configurations are different and their ratio
 depends on the time of calculation. To estimate the fusion probability
 in the reactions under discussion  we have used the values of the master equation
 solutions $P_K$ at  $t_{\rm fus}=6 \cdot 10^{-22}$s starting after capture
 to see the effect of the non-equilibrium stage of the charge and
 mass distributions between nuclei of DNS. The dependence of the neutron distribution
as a function of the charge number of the light fragment of the
dinuclear system formed in the $^{34}$S+$^{208}$Pb and
 $^{36}$S+$^{206}$Pb reactions are presented in Figs. \ref{NZ34S208Pb} and
 \ref{NZ36S206Pb}, respectively.

The probability of the yield of the quasifission fragment with the mass and charge
numbers, $A$ and $Z$, respectively,  after interaction time $t_{\rm int}$ of
DNS is estimated by
\begin{equation}\label{yield}
Y_{A,Z}(E^*_Z(A),\ell,t_{int})=\int\limits^{t_{\rm int}}_{0}P_{A,Z}(E^*_Z(A),\ell,t)
\Lambda^{\rm qf}_{A,Z}dt,
\end{equation}
where $P_{A,Z}^{}(E^*_{A,Z},\ell,t)$ is the probability of population of
the configuration $(Z,A)$ of the DNS at the given values of the excitation energy $E^*_Z(A)$,
 angular momentum $\ell$ and interaction time $t$. The interaction time $t_{\rm int}$ is
 calculated as the decay time of the DNS (details of calculation
 are presented in Ref. \cite{EPJA34}).
  The analysis of the yield and
 mass distribution of quasifission products is not subject of this work.
 This important and interesting research is the aim of our work in future
 to make a conclusion about fusion mechanism of heavy nuclei.

  In this work we use only
 mass and charge distributions $P_{A,Z}=P_Z(A,t)\times P_N(A,t)$ which are
 used to find most probable values of $N$ corresponding to the charge numbers
 $Z$ of the DNS fragments. The results of calculation
  of neutron distribution for the given proton number for
  the $^{34}$S+$^{208}$Pb and $^{36}$S+$^{206}$Pb reactions are presented in Fig(s).
  \ref{NZ34S208Pb} and \ref{NZ36S206Pb}, respectively. The numbers on the contours
  show probability of the proton and neutron distributions in the projectile-like
  fragments of the DNS formed at capture.

  The charge number $Z$ and corresponding mass number $A$ are
 used to calculate PES which allows us to calculate  the fusion probability $P_{\rm CN}$
 as a function of the mass and charge asymmetry  of the DNS nuclei.
 Therefore, the contributions to the complete
 fusion of different configurations are different and their ratio
 depends on the time of calculation.

\subsection{Fusion cross section}
\label{Fusion}

The partial fusion cross section is determined by the
product of capture cross section $\sigma^{(\ell)}_{cap}(E^*_{\rm
CN},\{\beta_i\})$ and the fusion probability $P_{CN}$ of DNS for
the various excitation energies
\cite{FGLM04,KKNMG15,FGRMN05,FGHMM09,ACNPV93} by the use of formula:
\begin{equation}
\label{sigmafus} \sigma^{(\ell)}_{fus}(E_{\rm
c.m.},\{\beta_i\})=P_{CN}(E_{\rm c.m.},\ell,\{\beta_i\})\sigma^{(\ell)}_{cap}(E_{\rm
c.m.},\{\beta_i\}),
\end{equation}
as the sum of contributions to complete fusion from the charge symmetric
 configuration $Z_{sym}$ of DNS up to configuration corresponding to the maximum value
 of the driving potential $Z_{max}$:
\begin{equation}
P_{CN}(E_{\rm c.m.},\ell,\{\beta_i\})=\sum\limits^{Z_{max}}_{Z_{sym}}
P_Z(E^*_Z,\ell)P^{(Z)}_{CN}(E^*_Z,\ell,\{\beta_i\}),
\label{PCNZ}
\end{equation}
where $E^*_Z$ is calculated by formula (\ref{Edns}) and the weight function $P_Z(E^*_Z,\ell)$
is the mass and charge distributions probability $P_Z(E^*_Z,\ell)$ in the DNS fragments
is determined by solution of the transport master equation (\ref{massdec});
the fusion probability $P^{(Z)}_{CN}(A)$ from the charge ($Z$) and mass ($A$) asymmetry
configuration of the DNS
 is calculated as the branching ratio $P^{(Z)}_{CN}(E^*_{Z},\ell;\{\alpha_i\})$
 of widths related to the overflowing over the
 quasifission barrier $B_{qf}(Z)$ at a given mass asymmetry,
 over the intrinsic barrier $B_{fus}(Z)$ on mass asymmetry axis
 to complete fusion and over  $B_{sym}(Z)$ in opposite direction
 to the symmetric configuration of the DNS:
\begin{equation}
 \label{Gammas} P^{(Z)}_{CN}\approx
 \frac{\Gamma_{fus}(Z)}{\Gamma_{qfiss}(Z)+ \Gamma_{fus}(Z)+\Gamma_{sym}(Z)}.
 \end{equation}
Here, the complete fusion process is considered as the evolution
of the DNS along the mass asymmetry axis overcoming
$B_{fus}(Z)$ (a saddle point between $Z=0$ and $Z=Z_P=16$) and
ending in  the region around $Z=0$ or $Z=Z_{CN}$ (fig.
\ref{driving}). The evolution of the DNS in the direction of the
symmetric configuration increases the number of events leading to
quasifission of more symmetric masses. This kind of channels are
taken into account by the term $\Gamma_{sym}(Z)$. One of the similar ways
was used in Ref.\cite{AdamPRC69}.
The complete fusion
can be presented by the formula of the  width of  usual fission
\cite{Siemens}:
\begin{equation}
\label{Gammai} \Gamma_{fus}(Z)=\frac{\rho_{\rm fus}(E^*_{Z})T_Z}{2\pi\rho(E^*_{Z})}
\left(1-\exp\frac{(B_{\rm fus}(Z)-E^*_{Z})}{T_Z}\right),
\end{equation}
where
 $\rho_{\rm fus}(E^*_{Z})=\rho(E^*_{Z}-B_{\rm fus}(Z))$;
  usually the value of the factor
$$\left(1-\exp\left[(B_i(Z)-E^*_{Z})/T_Z\right]\right)$$ in
 (\ref{Gammai}) is approximately equal to the unit. Inserting Eq. (\ref{Gammai}) in
(\ref{Gammas}), we obtain  the expression (\ref{Pcn}) used in our
calculations \cite{EPJA34}:
\begin{equation}
 \label{Pcn} P^{(Z)}_{CN}(E^*_{Z})=\frac{\rho_{\rm fus}(E^*_{Z})
}{\rho_{\rm fus}(E^*_{Z}) +
\rho_{\rm qfiss}(E^*_{Z})+\rho_{\rm sym}(E^*_{Z})}.
 \end{equation}
Putting the level density function of the Fermi system leads to formula of the
calculation of fusion probability for the given values of the DNS excitation energy $E^*_Z$
and angular momentum $L$ from its charge asymmetry $Z$:
\begin{equation}
 \label{Pcn} P^{(Z)}_{CN}(E^*_Z)=\frac{e^{-B_{\rm fus}^{*(Z)}/T_Z}}{e^{-B_{\rm fus}^{*(Z)}/T_Z} +
e^{-B_{\rm qfiss}^{*(Z)}/T_Z}+e^{-B_{\rm sym}^{*(Z)}/T_Z}}.
 \end{equation}
The fusion cross section  is calculated by summarizing contributions
of all partial waves (angular momentum):
\begin{equation}
\sigma_{\rm fus}(E_{\rm c.m.})=\sum\limits^{\ell=\ell_d}_{\ell=0}
\langle\sigma^{(\ell)}_{\rm fus}(E_{\rm c.m.})\rangle
\label{sigmafus}
\end{equation}
The averaged value of the partial fusion cross section is calculated
by the same method as in the case of partial capture cross section:
\begin{eqnarray}
\langle \sigma^{(\ell)}_{\rm fus} (E_{c.m}) \rangle &=& \int^{\beta_{2+}}_{-\beta_{2+}}
 \int^{\beta_{3-}}_{-\beta_{3-}} \sigma^{(\ell)}_{\rm fus}(E_{\rm c.m}, \beta_2, \beta_3) \nonumber\\
&\times & g(\beta_2, \beta_3) d\beta_2 d\beta_3.
\label{avfus}
\end{eqnarray}

 The evaporation residue cross sections after emission 2 and 3 neutrons
 is calculated by the use of $\langle\sigma^{(\ell)}_{\rm fus} (E_{c.m})\rangle$
 to take into account  the dependence of the fission barrier on the angular momentum as in Ref.
 \cite{GMPhysRevC86}.

\section{Results of calculations}
\label{results}
\subsection{Capture of nuclei}\label{results1}

The trajectories of the relative motion for reactions $^{34}$S+ $^{208}$Pb and
$^{36}$S+$^{206}$Pb at collision energies around the Cou-lomb barrier
have been calculated by solving  the equations of motion
(\ref{maineq1})-(\ref{maineq5}) for the relative distance and
velocity. The bifurcation of the collision trajectories on the
deep-inelastic collision and capture as a function of orbital
angular momentum ($\ell$) at the given collision energy  $E_{\rm c.m.}$
  is calculated by the use of the friction
coefficient which is determined by the particle-hole excitation of the
nucleons in nuclei and nucleon exchange between them \cite{NKO16}.
At the sub-barrier energies the probability of capture is determined
by the WKB approximation.

It is seen that the
trajectory with the orbital angular momentum $\ell=0$ leads to capture
because the relative kinetic energy is enough to overcome barrier at
$E_{c.m.}=146.41$ MeV. The
rotational energy increases by rising $L$ and now the total energy is
not enough to overcome the barrier of the interaction potential, therefore,
starting from $\ell=50$  we observe the deep inelastic collisions only for this
reaction.

As a result of the capture, we have the DNS evolved by nucleon
transfer between the nuclei to reach the equilibrium mass  and charge
distribution which is determined by peculiarities of PES and
level densities of the single-particle states of protons and neutrons
in interacting nuclei. One of final states of evolution is a formation of the compound
 nucleus after the transfering  all nucleons from the light fragment to the heavy fragment.
 The alternative states of the evolving DNS are its decay into
 two fragments in dependence on the height of $B_{\rm qf}$ and
 the DNS excitation energy.

The results obtained for the charge and mass distributions by solving
master equations (\ref{massdec}) show that initial values of the ratio of
the neutron and proton numbers in colliding nuclei are important
in the calculation of the fusion probability.

\begin{figure}
\resizebox{0.65\textwidth}{!}{\includegraphics{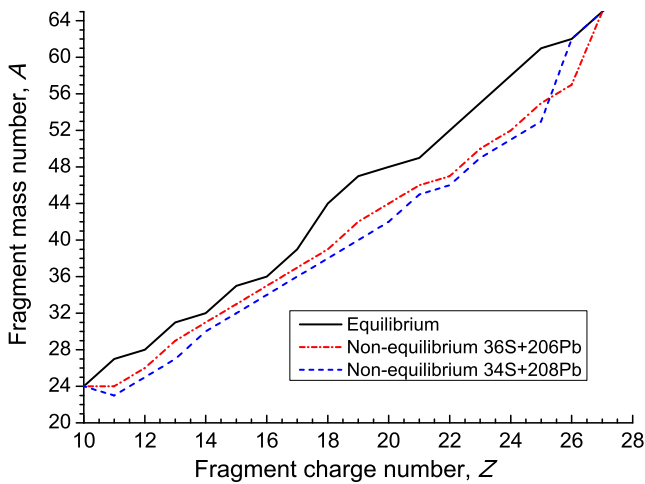}}
\vspace*{-3.0cm} \caption{(Color online) The mass number in
the projectile nucleus as a function of its proton number
calculated for $^{36}$S+$^{206}$Pb (dot-dashed line) and $^{34}$S+$^{208}$Pb
(dashed line) reactions for non-equilibrium initial stage of the DNS evolution.
The equilibrium distribution of neutrons between fragments
 corresponds to the minimum values of the PES as a function of mass
 numbers one of the DNS fragments (solid line).} \label{NZratio}
\end{figure}

 It is obvious from Fig.~\ref{NZratio} that the projectile-like
 fragments of the DNS formed in the $^{36}$S+$^{206}$Pb reaction
 are more neutron rich in comparison with the ones of the  $^{34}$S+$^{208}$Pb
 reaction. As a result the fusion probability is larger in the first reaction.
 Neutron numbers $N$ corresponding to the given charge numbers presented in
 Fig.~\ref{NZratio} are found from the analysis of the parallel solutions of the
 transport-master equations (\ref{massdec}): the neutron number  $N$ corresponding
 to the maximum value of the neutron distribution function  $P_N(A,t), (K=N)$
 for the given $Z$  is used in calculation of PES.
The equilibrium distribution of neutrons between fragments
 corresponds to the minimum values of the PES as a function of mass
 numbers one of the DNS fragments (solid line in Fig. \ref{NZratio}).

 So, the fusion probability of the DNS nuclei is determined by
 the intrinsic fusion barrier ($B^{\ast}_{fus}$) and
 quasifission barrier ($B_{qf}$) which are functions
 of the proton and neutron numbers (see Ref. \cite{NFTA05}).

This result has been obtained from the neutron distributions in the
 light fragment of DNS  as a function of its charge number at interaction
time $t_{\rm int}=6\cdot 10^{-22}$ s after capture (see Figs.
\ref{NZ34S208Pb} and \ref{NZ36S206Pb}).
The time preceding to capture from the beginning the dissipation
of the relative energy is about $4\cdot10^{-22}$---$6\cdot10^{-22}$ s as
function of the values of $E_{\rm c.m.}$ and $\ell$.
\begin{figure}
\vspace*{-0.55cm}
\hspace*{-0.45cm}
\resizebox{0.525\textwidth}{!}{\includegraphics{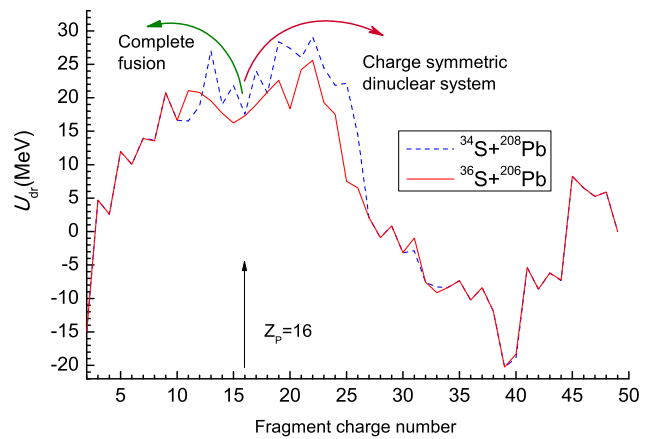}}
\vspace*{-1.25cm} \caption{(Color online) Driving potential calculated
for the compound nuclei $^{242}$Cf reaction as a function of the
fragment charge and mass number.} \label{driving}
\end{figure}
It can be seen from Fig. \ref{driving} that the driving potential
(blue dashed line) calculated for the $^{34}$S+$^{208}$Pb reaction
increases abruptly for the fragment with charge number $Z=13$.
 The value of the driving potential corresponding to the
entrance channel $Z=16$ is lower than its maximum value at $Z=13$ in the fusion
direction $Z\rightarrow0$.
The increase of the hindrance to complete fusion in the $^{34}$S+$^{208}$Pb reaction
in comparison with the  $^{36}$S+$^{206}$Pb reaction is seen from the comparison
of PES in Figs. \ref{PES34S} and \ref{PES36S} which are calculated as functions of  the
intercentre distance between nuclei and their charge-mass asymmetry.
The difference between the two PES(s) in Figs. \ref{PES34S} and \ref{PES36S}
appears due to the use of the different mass numbers obtained in the solution of
the Eqs. (\ref{massdec}) by the different initial neutron numbers.
As it is seen from Fig. \ref{PES34S}  the potential surface has higher bump
corresponding to the intrinsic fusion barrier, $B^{\ast}_{fus}$, in the region
$Z=13$ and $R=13.5$ fm. This bump appears as the hindrance in complete
fusion  in the case of the $^{34}$S + $^{208}$Pb reaction. This bump is
significantly higher than the one on the potential energy surface
presented in Fig. \ref{PES36S} for the $^{36}$S + $^{206}$Pb reaction.
  The hindrance to the DNS evolution in the direction
of the symmetric charge distributions is determined by the barrier
$B^{\ast}_{sym}$ which is determined in a similar way to the case of
$B^{\ast}_{fus}$ but the maximum value of the driving potential
from symmetric charge region
($U_{dr}(Z^{sym}_{max},A^{sym}_{max},\ell)$) is used.

\begin{figure}
\vspace*{+2.46cm}
\hspace*{-0.25cm}
\resizebox{0.50\textwidth}{!}{\includegraphics{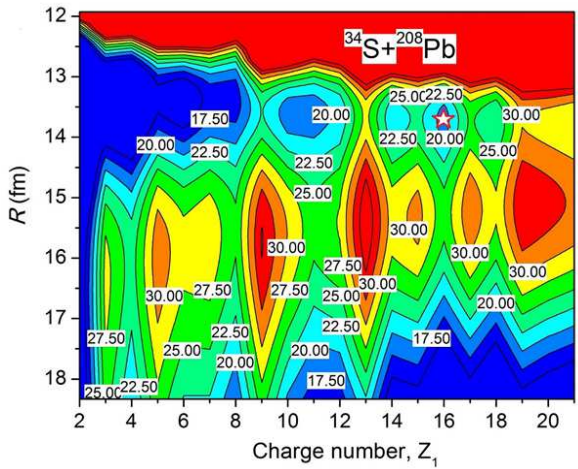}}
\vspace*{-3.25cm} \caption{(Color online) Contour map of the PES calculated for the
 $^{34}$S+$^{208}$Pb reaction with the non-equilibrium distribution of neutrons
 between the DNS fragments as a function of the radial distance between
 their mass centres and charge numbers. The point of PES corresponding
 to the initial charge and mass numbers is shown with a star.} \label{PES34S}
\end{figure}

\begin{figure}
\vspace*{+2.46cm}
\resizebox{0.55\textwidth}{!}{\includegraphics{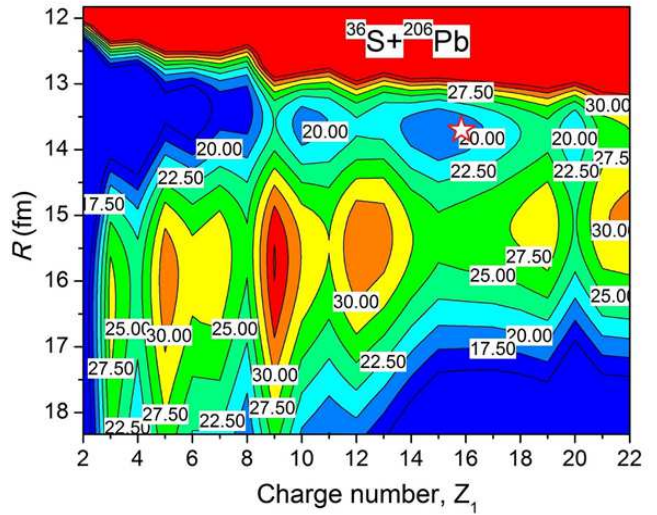}}
\vspace*{-3.25cm} \caption{(Color online) The same as in Fig. \ref{PES34S}
 but  for the $^{36}$S+$^{206}$Pb reaction.} \label{PES36S}
\end{figure}

\begin{figure}
\hspace*{-0.25cm}
\resizebox{0.52\textwidth}{!}{\includegraphics{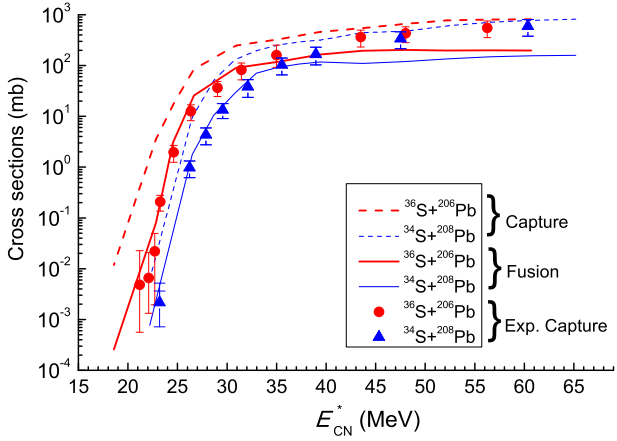}}
\vspace*{-1.5cm} \caption{(Color online) Capture and complete fusion cross sections
calculated for $^{36}$S+$^{206}$Pb (thick and thin dashed curves) and $^{34}$S+$^{208}$Pb reactions
(thick and thin solid curves) as a function of the CN excitation energy are
compared  with the experimental data \cite{KNHAB12}.} \label{fuscross}
\end{figure}

In Fig.~\ref{fuscross} the capture and complete fusion cross sections are compared
with the experimental data.
The excitation energy  of the compound nucleus $E^*_{\rm CN}= E_{\rm c.m.}+Q_{gg}$
corresponding to the collision energy in the center of mass system $E_{\rm c.m.}$
has been used for the convenience of comparison of the
corresponding experimental and theoretical cross sections of these reactions.
It is clearly seen in Fig.~\ref{fuscross} that  the excitation fusion function of the
$^{36}$S+$^{206}$Pb is much higher than the one of the $^{34}$S+$^{208}$Pb reaction
 in the energy region $E^*_{\rm CN}$=24--35 MeV, which corresponds to experimental
 results \cite{KNHAB12}.

\begin{figure}
\vspace*{0.0cm}
\resizebox{0.65\textwidth}{!}{\includegraphics{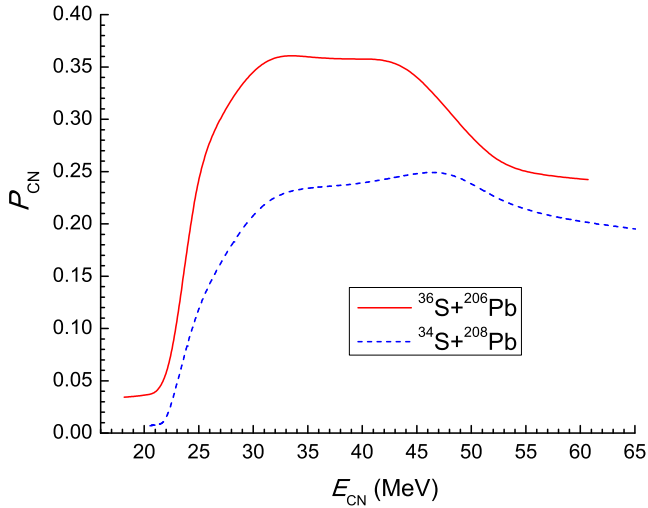}}
\vspace*{-3.0cm} \caption{(Color online) Fusion probability $P_{\rm CN}$
 calculated for the $^{36}$S+$^{206}$Pb (red solid line) and $^{34}$S+$^{208}$Pb
 (blue dashed line) reactions.}
\label{PCN36S34S}
\end{figure}

 The lower threshold energy  $E^{*min}_{\rm CN}$ of the fusion excitation function is
 determined by the height of the Coulomb barrier in the entrance channel
 and reaction balance energy $Q_{gg}$. The large negative values of $Q_{gg}$
 decrease the value of  $E^{*min}_{\rm CN}$ \cite{KKNMG15}.
 The $Q_{gg}$-values are equal to -113.79 and -111.02 MeV for the
  $^{36}$S+$^{206}$Pb and $^{34}$S+$^{208}$Pb reactions, respectively.
  As it was discussed above as well as according to Fig. \ref{potenell}, the Coulomb barrier
  of the $^{36}$S+$^{206}$Pb reaction is lower than the one of the $^{34}$S+$^{208}$Pb
  reaction. Consequently, the threshold energy  $E^{*min}_{\rm CN}$ for the first
  reaction is significantly lower than the one for the second reaction. Therefore,
  the condition to increase the evaporation residue cross sections has been revealed.

\begin{figure}
\vspace*{0.0cm}
\resizebox{0.65\textwidth}{!}{\includegraphics{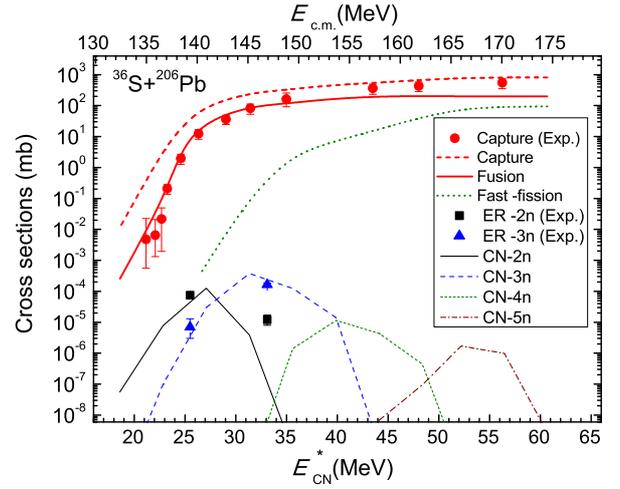}}
\vspace*{-2.8cm} \caption{(Color online) The theoretical values of capture
(red thick dashed curve), complete fusion (solid thick curve) and ER
  (thin solid-2n, thin dashed-3n, thin dotted-4n and thin dot-dashed-5n channels)
  cross sections are compared with the experimental values of the capture (red circles)
  and ER (black squares-2n and red triangles-3n channels)
  cross sections of the $^{36}$S+$^{206}$Pb \cite{KNHAB12}.}
\label{ERS36Pb206}
\end{figure}

Comparison of the fusion probabilities $P_{\rm CN}$ calculated for the
 $^{36}$S+$^{206}$Pb and $^{34}$S+$^{208}$Pb reactions is presented in
 Fig. \ref{PCN36S34S}. It is seen that the complete fusion probability
 of  the $^{36}$S+$^{206}$Pb reaction is about one and half times larger
 than that of the $^{34}$S+$^{208}$Pb reaction.
 This factor  leads to the increase additionally in the difference  between
 ER cross sections of these reactions.
 The large value of the capture cross section of the $^{36}$S+$^{206}$Pb
 reaction causes the larger value of the ER
 cross section of this reaction in comparison with the $^{34}$S+$^{208}$Pb reaction.

\begin{figure}
\vspace*{0.0cm}
\resizebox{0.65\textwidth}{!}{\includegraphics{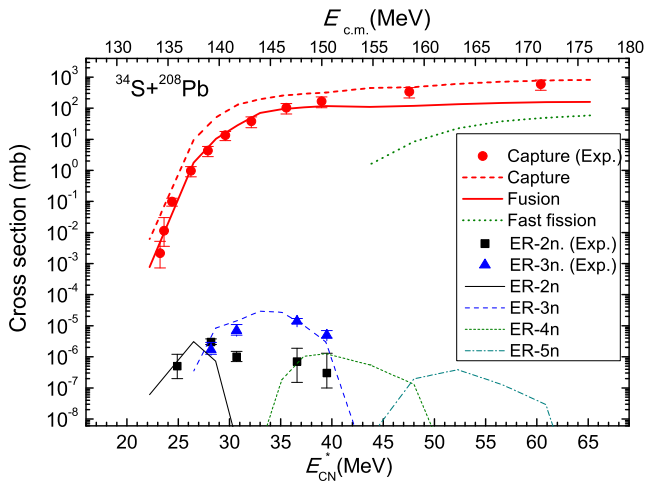}}
\vspace*{-2.8cm} \caption{(Color online) The same as in Fig. \ref{ERS36Pb206}
but for the  $^{34}$S+$^{208}$Pb reaction \cite{KNHAB12}.}
\label{ERS34Pb208}
\end{figure}

 The theoretical values of capture, complete fusion and ER
 cross sections are compared with the experimental values of the capture and
 ER cross sections of the $^{36}$S+$^{206}$Pb and
   $^{34}$S+$^{208}$Pb reactions in Figs. \ref{ERS36Pb206} and \ref{ERS34Pb208},
   respectively.  The partial fusion cross sections are used as the input data
   in calculations of the ER cross sections  by the advanced statistical model
   \cite{GMPhysRevC86}.
 It is seen from these figures that the theoretical results for the 3n-evaporation channel
 are in good agreement with the experimental data while the theoretical curve obtained
 for the 2n-evaporation channel is in good agreement with the data up to energies
 $E^*_{\rm CN}$=30 MeV and 28 MeV for the  the $^{36}$S+$^{206}$Pb and
   $^{34}$S+$^{208}$Pb reactions, respectively. To reach an agreement for the 2n-evaporation
    channel it seems an additional assumption made in the calculation of the
    advanced statistical model.

\section{Conclusion}
\label{conclusion}
The difference between observed cross sections of the
ER  of the $^{34}$S+$^{208}$Pb and $^{36}$S+$^{206}$Pb reactions
formed in the 2n and 3n channels has been explained by two reasons
related to the entrance channel characteristics of these reactions.
The first reason is the difference in the sizes and position of the potential
  wells of the nucleus-nucleus interaction calculated for these reactions.
  The presence of two extra neutrons in isotope $^{36}$S and
  projectile-like fragments makes the potential well  deeper and lower
  for the $^{36}$S+$^{206}$Pb reaction.
  Therefore, the capture cross section for this reaction is larger than the one of the
$^{34}$S+$^{208}$Pb reaction, {\it i.e.} more the number of DNS being
to be transformed into compound nucleus is formed in the $^{36}$S+$^{206}$Pb reaction.
The second reason is the difference in the heights of the intrinsic fusion
barrier $B^*_{\rm fus}$ appearing on the fusion trajectory by nucleon transfer between
nuclei of the DNS formed after capture. The value of $B^*_{\rm fus}$
 calculated  for the $^{34}$S+$^{208}$Pb reaction is higher than the one obtained for the
   $^{36}$S+$^{206}$Pb  reaction. This fact is caused by the difference between
   the $N/Z$-ratios in the light fragments of the DNS formed during the capture in
   these reactions.
The  $N/Z$-ratio has been found by solution of the transport master equations
 for the proton and neutron distributions between fragments of the DNS
 formed at capture with the different initial neutron numbers
 $N=18$ and $N=20$ for the reactions with the $^{34}$S and $^{36}$S, respectively.

 These reasons are related to the
 largest of the $N/Z$-ratio in the projectile-like
 fragments in the  $^{36}$S+$^{206}$Pb reaction at the initial non-equilibrium stage
 of the interaction of the DNS fragments. This is seen from the comparison of the shape
  of the driving potentials and landscape of PES, which are calculated for these two reactions.
  In the  DNS with the neutron-rich projectile-like fragments formed in the
   $^{36}$S+$^{206}$Pb reaction the intrinsic fusion barrier is lower.
  Due to these two consequences, the use of the neutron rich isotope $^{36}$S makes
    the ER cross section larger in  the $^{36}$S+$^{206}$Pb reaction
   at the de-excitation of compound nucleus in comparison with the ones of
   the $^{34}$S+$^{208}$Pb reaction. A larger hindrance to complete fusion in
    the reaction with $^{34}$S  can be observed from the analysis of the yield of the
     projectile-like capture products observed in both reactions under discussion.
    The intense yield of the projectile-like capture products decreases the number events
     going to complete fusion which produces fusion-fission products and evaporation
      residues after emission neutrons and light charged particles.

\begin{acknowledgement}
This paper has been done in the framework of the Project F2-FA-F115 of the Committee for coordination
science and technology development under the Cabinet of Ministers of Uzbekistan.
The authors A.K.N. and B.M.K. thank to the committee for their financial support.
A.K.N. thanks to the Russian Foundation for Basic Research for the partial support.
One of the authors B.M.K. is grateful to Rare Isotope Science Project of Institute for Basic Science
for the support of his staying in Daejeon.
The work of K.K and Y.K is supported by the
Rare Isotope Science Project of Institute for Basic Science
funded by the Ministry of Science and ICT and National
Research Foundation of Korea (2013M7A1A1075764).
\end{acknowledgement}

\end{document}